%% file: arxiv_submission.tex
\documentclass[trackchanges]{aastex7}

\newcommand{\chandra}{{\em Chandra}}
\newcommand{\XRISM}{{\em XRISM}}

\usepackage{soul}
\usepackage{amsmath}

\begin{document}

\title{A Yin-Yang Galaxy Cluster Merger in Abell 1914 Revealed by \XRISM{}\\
}

\author[orcid=0000-0002-7726-4202,gname='Annie', sname='Heinrich']{Annie Heinrich}
\affiliation{University of Chicago, Department of Astronomy \& Astrophysics, Chicago, IL 60637, USA}
\email[show]{amheinrich@uchicago.edu}  

\author[orcid=0000-0001-5888-7052,gname=Congyao,sname=Zhang]{Congyao Zhang}
\affiliation{Masaryk University, Department of Theoretical Physics \& Astrophysics, Brno 61137, Czechia}
\affiliation{University of Chicago, Department of Astronomy \& Astrophysics, Chicago, IL 60637, USA}
\email{placeholder}

\author[orcid=0000-0001-7630-8085,gname=Irina, sname=Zhuravleva]{Irina Zhuravleva} 
\affiliation{University of Chicago, Department of Astronomy \& Astrophysics, Chicago, IL 60637, USA}
\email{zhuravleva@uchicago.edu}

\author[orcid=0000-0002-0322-884X,gname=Eugene,sname=Churazov]{Eugene Churazov}
\affiliation{Max Planck Institute for Astrophysics, Garching, Germany}
\email{placeholder}

\author[orcid=0000-0003-3537-3491,gname=Hannah, sname=McCall]{Hannah McCall} 
\affiliation{University of Chicago, Department of Astronomy \& Astrophysics, Chicago, IL 60637, USA}
\email{placeholder}

\author[orcid=0000-0002-0587-1660,gname=Reinout J., sname=van Weeren]{Reinout J. van Weeren} 
\affiliation{Leiden Observatory, Leiden University, 2300 RA Leiden, The Netherlands}
\email{placeholder}

\author[orcid=0000-0002-9478-1682,gname=William R., sname=Forman]{William R. Forman} 
\affiliation{Center for Astrophysics, Harvard and Smithsonian, Cambridge, MA 02138, USA}
\email{placeholder}

\begin{abstract}

Hierarchical mergers of galaxy clusters play a key role in converting gravitational energy into thermal and kinetic energy in the local universe.
Understanding this process requires the reconstruction of cluster merger geometry, with careful consideration of projection effects.
With its unprecedented spectral resolution, \XRISM{} enables the disentanglement of merging cluster components along the line-of-sight via X-rays for the first time.
In this letter, we focus on the massive cluster A1914, a puzzling case wherein the galaxy and dark matter distribution appear to be in tension with the X-ray morphology.
We present \XRISM{} observations of A1914 focusing on the velocity structure of the intracluster medium (ICM).
The \textit{Resolve} full-array spectrum requires two merging components along the line-of-sight, with bulk velocities offset by $\sim 1000$ km/s and velocity dispersions of $\sim 200$ km/s.
The sub-array maps of flux ratios, bulk velocity, and velocity dispersion show the two components are offset and overlapping in the plane of the sky, consistent with a major (mass ratio $\sim 3$), near line-of-sight merger with a pericenter distance of $\sim 200$ kpc. 
We conclude that the two subclusters create an overlapping spiral pattern, referred to as a ``yin-yang'' merger.
This scenario is further supported by tailored hydrodynamical simulations of the A1914 merger, demonstrating that this type of merger can broadly reproduce the observed X-ray morphology, gas temperature map, gas velocity maps, dark matter distribution, and galaxy velocities. 
This work demonstrates the power of high-resolution X-ray spectroscopy, provided by \XRISM{}, to resolve complex cluster merger geometries.

\end{abstract}


\section{Introduction}
\label{sec:intro}
Hierarchical mergers between massive galaxy clusters are the most energetic events in the modern universe, converting up to $\sim 10^{65}$ ergs of gravitational potential energy to kinetic energy of dark matter (DM), galaxies, and the intracluster medium (ICM) \citep[see, e.g., a recent review by][and references therein]{zuhone2022}, the latter of which is an important heating mechanism in the local universe.
Constraining the geometry of these mergers is crucial to understanding energy conversion and distribution in these large-scale structures.
Although X-ray and optical observations have allowed us to study many of these violent events \citep[e.g.,][]{briel_observation_1992,markevitch_textbook_2002,kempner_chandra_2003,wang_merging_2016}, reconstructing their underlying geometries remains challenging.
The recently launched \XRISM{} satellite \citep{xrismcollab_firstlight}, with its \textit{Resolve} microcalorimeter \citep{ishisaki2022_reffix}, can deliver high-resolution spectra of the hot ICM.
This is the most powerful existing method to directly measure ICM kinematics in merging clusters, providing more accurate measurements than the kinetic Sunyaev-Zel'dovich (SZ) effect \citep[e.g.,][]{adam2017} or X-ray CCD spectroscopy \citep{gatuzz2023,gatuzz2024} and making it possible to disentangle multiple ICM components along the line of sight for the first time.
\XRISM{} can therefore provide crucial insight into the geometry of cluster mergers.
Thus far, only a handful of merging clusters have been studied with \XRISM{}: Coma \citep{xrismcollab_coma}, A2319 \citep{xrismcollab_a2319}, and Ophiuchus \citep{fujita2025}.

In this work, we present XRISM/Resolve observations of A1914, a $z \approx 0.168$ \citep[][]{barrena2013} cluster undergoing a major merger, focusing on measuring gas kinematic properties.
A1914 is relatively luminous and is at a higher redshift than most other clusters observed by \XRISM{} so far.
Weak lensing measurements indicate this cluster has a double-peaked mass distribution, elongated along the northeast-southwest axis \citep[e.g.,][]{mandal2019,ami2012,okabe2008}.
This disturbed DM morphology differs from the X-ray morphology, which is approximately spherical at large radii \citep{mann2012}, has a cool, bright, low entropy region just east of the cluster center \citep{botteon2018}, and three identified shocks \citep{rahaman2022}.
A1914 also contains diffuse radio emission in the form of a filamentary radio phoenix in the east of the cluster, and a radio halo in the west \citep[see, e.g.,][and references therein]{mandal2019,bacchi2003}.
The phoenix is approximately co-spatial with the cool region, while the halo is not obviously associated with any X-ray features.

Previous studies of this cluster \citep[e.g.,][]{rahaman2022, botteon2018} suggest the cool, bright region could be explained by a head-on, ``Bullet-like'' merger, perhaps in combination with another merger that could explain the NE-SW mass elongation.
In this scenario, the cool region is a remnant cool core in the process of being ram-pressure-stripped.
As this subcluster passes through the main cluster, a bow shock develops ahead of the cool-core remnant, and a cold front is visible around the remnant \citep{markevitch_textbook_2002}.
Though a shock front is detected ``ahead'' of the bright region (to the southeast), no cold front is detected.
The positioning of the bright region and bow shock would additionally imply an east-west merging axis, inconsistent with the merger axis implied by the DM distribution and positions of the brightest cluster galaxies (BCGs) \citep[e.g.,][]{barrena2013}.

In this work, we test this merger geometry using ICM gas velocities measured with \XRISM{}.
We find the velocity structure of A1914 inconsistent with a head-on, Bullet-like merger.
Instead, we propose a new configuration, where a major merger is occurring with a large impact parameter along the line of sight.
In this scenario, the two subclusters are partially overlapping in projection, and their respective gas components create a ``yin-yang''-like morphology.

This paper is organized as follows: Section \ref{sec:methods} describes the observations, production of spectra, and modeling process; Section \ref{sec:results} presents the measured ICM properties using three different modeling strategies and a merger geometry with supporting simulations to explain the observed velocity structure.
For this work, we assume a $\Lambda$CDM cosmology with $h=0.7$ and $\Omega_m=0.3$, wherein at the redshift of A1914, 1 arcminute corresponds to $\simeq 175$ kpc.
Uncertainties are $1 \sigma$ unless stated otherwise.

\section{Methods}

\label{sec:methods}
\subsection{\XRISM{} observations and data reduction}
A1914 was observed by \XRISM{} in January 2025 (ObsID 201093010).
This observation was planned as part of an effort to observationally calibrate the relationship between ICM gas velocities and density fluctuations \citep[e.g.,][]{zhuravleva2014,heinrich2024}.
This observation was pointed at the cluster center (J2000: RA=216.5003$^\circ$, DEC=37.8264$^\circ$)  and was reprocessed using \textsc{heasoft} version 6.35.1 with \textit{Resolve} \textsc{CalDB} version 20250315.
The \textit{Resolve} field-of-view (FOV) covers a $\simeq$525$\times$525 kpc region \citep[$\sim 0.75 R_\mathrm{2500}$, ][Fig. \ref{fig:fullfov}]{rahaman2022}.
We screened the produced events list for High-resolution primary (Hp) events as described in the \XRISM\ Quick Start Guide \citep[for details, see][]{xrismcollab_firstlight}, yielding data with a spectral resolution of $\simeq 4.5$ eV.
After screening, the exposure time of the observation was $\simeq 112$ ks.
Pixel 27, known to have an unstable gain, was removed from our data. 

We employ three spatial binning strategies (see Section \ref{sec:results}) to fully investigate the ICM velocity structure of A1914. 
For each binning strategy, source spectra were extracted from one or more detector regions, large-size redistribution matrix files (RMFs) were produced with the \texttt{rslmkrmf} script for each region, and auxiliary response files (ARFs) were created with the \texttt{xaarfgen} tools. 
Non-X-ray background (NXB) spectra were made using \texttt{rslnxbgen}, taking data from \textit{Resolve}'s night-Earth database version 2.
The NXB event lists are filtered equivalently to the on-source spectra, and use a diagonal RMF with no ARF for fitting.
NXB spectra are extracted from the entire FOV, fit with an empirical model, and then scaled to the appropriate number of pixels in each detector region (see Section \ref{sec:nxb} below).
Source and background spectra were grouped with the \texttt{ftgrouppha} script, such that each energy bin contained at least one count.
{Each spectrum or set of spectra are then modeled with a combination of the NXB model and one or more ICM components, in the 2.5-11 keV band, using Cash statistics \citep[C-stat,][]{cash1979}.}

To create ARFs, we use an exposure-corrected \chandra{} ACIS image of A1914 in the 2-8 keV band (described in Section \ref{sec:chandra}), which most closely matches the nominal \textit{Resolve} band.
This image is masked such that it only contains a $3'$ radius circle around the center of the \textit{Resolve} FOV.
{This region was chosen to cover an area of the sky from which nearly all of the photons detected by \textit{Resolve} originated.}
For the majority of our binning strategies, which do not account for spatial-spectral mixing (SSM), we use the entire image to generate ARFs.
For the final binning strategy (described in Section \ref{sec:results}), we account for SSM by selecting sky regions in the \chandra{} image around each detector region.
{The central region covers the region matching the selected \textit{Resolve} pixels in that region, while the north and south regions each cover the rest of their respective halves of the $3'$ image.}
We then extract one ARF from each sky region for each detector region.
In the case of three detector/sky regions, we therefore create nine ARFs to account for SSM.
For more information on SSM, see, e.g., \citet{xrismcollab_A2029_out}.

\subsubsection{Spectral modeling: intracluster medium}
We model each ICM component as an emission spectrum from optically-thin plasma in collisional ionization equilibrium using AtomDB version 3.1.3. 
For this, we used the \texttt{bvapec} model in xspec along with \texttt{tbabs} to account for foreground absorption.
The hydrogen column density along the line of sight was fixed to $1.06 \cdot 10^{20}$ cm$^{-2}$ \citep{willingale2013}.
We use the solar abundance table \texttt{lpgs} \citep[][]{lodders2009}.
He and C abundances are fixed to solar values, while all other metallicities other than Ni are tied to Fe. 
Thus, the free parameters are gas temperature, Fe and Ni metallicities, redshift, velocity dispersion, and normalization.
{When modeling with multiple ICM components, each is modeled as a separate absorbed \texttt{bvapec} component; however, the data quality requires abundances to be tied between the components.
Temperature, redshift, velocity dispersion, and normalization are independent in each component.}

\subsubsection{Spectral modeling: NXB}
\label{sec:nxb}
The NXB spectrum is fit separately to the on-source spectra with an empirically-derived model \citep{Kilbourne2018} that includes a continuum component and 12 detector emission lines.
As advised, the continuum component is modeled as a power-law and fit while the emission line normalizations are fixed.
Subsequently, the continuum is frozen and line normalizations are freed.
This NXB model is frozen and scaled to the number of pixels in each detector region, then included in each subsequent on-source spectral fitting. 
In the Full-FOV spectrum, the background count rate is $\approx 6\cdot 10^{-4}$ counts s$^{-1}$ keV$^{-1}$ at 6 keV, a factor of $\sim 50$ lower than the observed count rate.

\begin{figure}
    \centering
    \includegraphics[width=\linewidth]{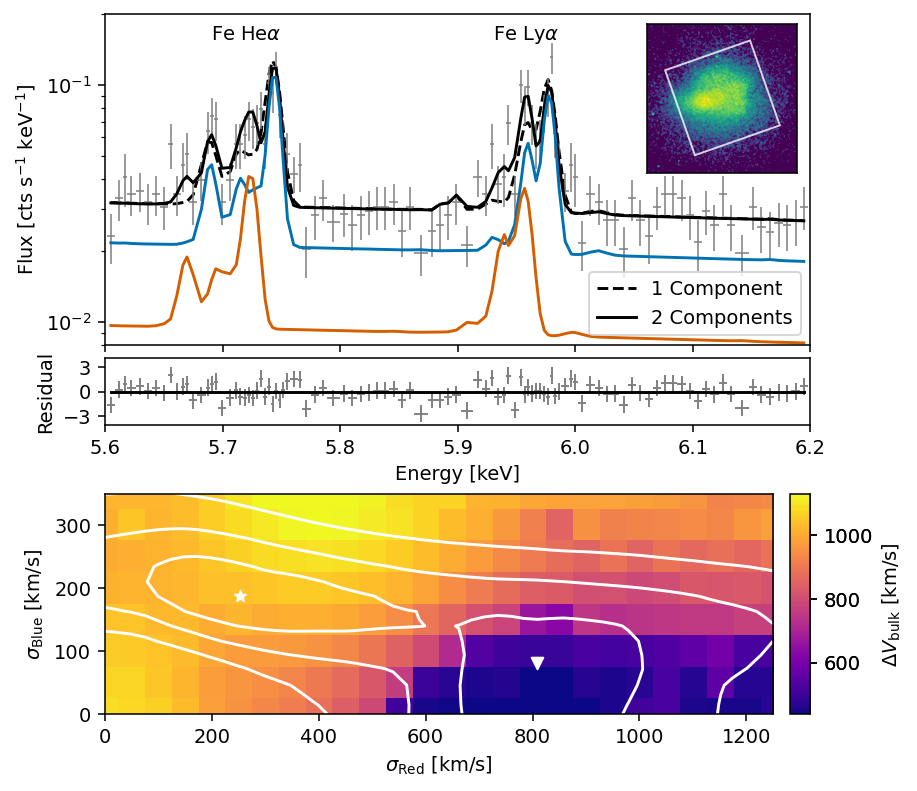}
    \caption{\textbf{Top:} 5.6-6.2 keV \textit{Resolve} spectrum of A1914 from the full FOV, showing Fe He$\alpha$ and Ly$\alpha$ lines. The spectrum is binned such that each energy bin is detected at $4 \sigma$ significance. The best-fit one-component model is shown as a dashed black line.
    The best-fit two-component model is shown in solid black, with individual ICM components shown in blue and red. 
    Residuals from the two-component model, calculated as (data-model)/error, are plotted below the spectrum.
    \textbf{Inset:} 2-8 keV \chandra{} image of A1914, with the \textit{Resolve} FOV (a simplified $3' \times 3'$ square)  overlaid in white.
    \textbf{Bottom:} Best-fit bulk velocity separation between the red and blue ICM components, plotted as color against the velocity dispersions of the two components. 1/2/3 $\sigma$ constraints on $\sigma_\mathrm{red}$ and $\sigma_\mathrm{blue}$ are plotted as solid white contours. Global/local minima are labeled with a star/triangle marker, respectively. }
    \label{fig:fullfov}
\end{figure}

\subsection{\chandra{} analysis}
\label{sec:chandra}
\begin{figure}
    \includegraphics[width=\textwidth]{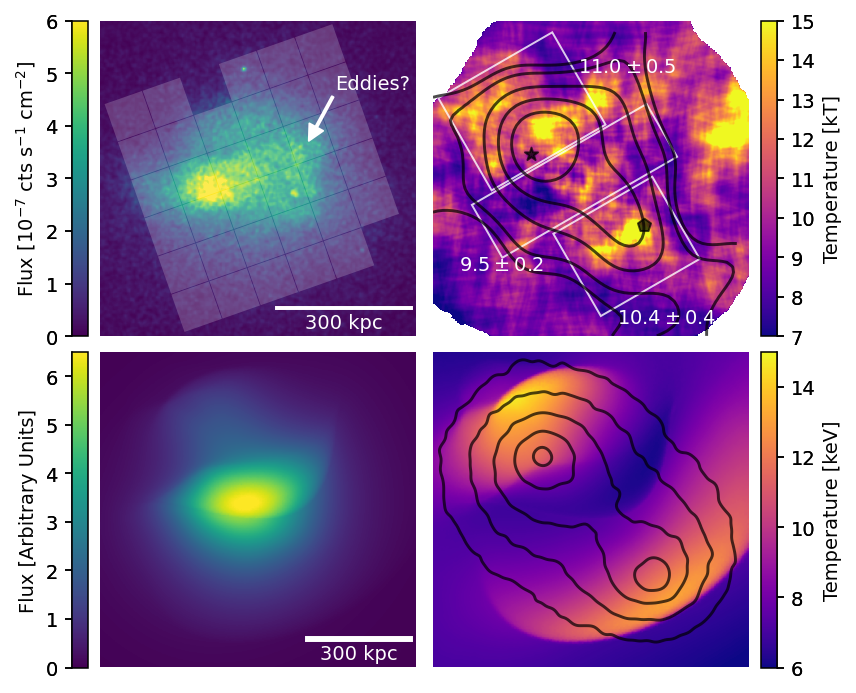}
    \caption{
    \textbf{Top:} 0.5-8 keV X-ray surface brightness map and temperature map from \chandra{}. In the surface brightness map, \XRISM{} pixels are overlaid in transparent white. In the temperature map, weak lensing contours are shown in black \citep{mandal2019}, and regions from which ICM temperatures are measured are shown as white boxes, with their respective temperatures labelled. We also plot the locations of the BCGs as a black star/pentagon.
    \textbf{Bottom:} Simulated X-ray surface brightness and emissivity-weighted temperature map, projected along the line of sight. DM density contours are plotted in black. 
    } 
    \label{fig:chandra}
\end{figure}

We reprocess \chandra{} ObsIDs 3593, 18252, 20023, 20024, 20025, and 20026 using the standard procedure described by \cite{vikhlinin2005}, with \textsc{ciao} version 4.14 and \textsc{caldb} version 4.9.7. 
Flares are filtered, and background event lists are produced accounting for blanksky and readout backgrounds.
We create mosaic images of A1914 in the 2-8 and 0.5-8 keV bands using these observations, which are presented in Figures \ref{fig:fullfov} and \ref{fig:chandra}. 

{To calculate the projected gas temperature map, we employ the method described in \cite{churazov1996,churazov2003}.  
The method relies on using template spectra of optically thin plasma at reference temperatures convolved with the \textsc{ACIS-I} energy response. 
The spectrum in each region is presented as a linear combination of these spectra with some coefficients that are the two free parameters of the model. 
Once the maps of the two coefficients are found, they are adaptively smoothed and combined to derive the value of the temperature. 
For A1914, the reference temperatures of 6, 10, and 14 keV were selected to encompass the full range expected in A1914's ICM.
Figure \ref{fig:chandra} includes the resulting temperature map. 
The map is binned by adaptively choosing regions containing $\sim 2000$ counts. 
It shows a bulk gas temperature of $\sim 9 $ keV, with several distinct hotter clumps reaching up to 14 keV.
Overall, the temperature distribution and its substructure are in good agreement with earlier studies \citep[e.g.,][]{govoni2004, botteon2018, rahaman2022}.}

\section{Results and Discussion}
\label{sec:results}
\subsection{Detection of two velocity components}
We initially extract a spectrum from the full \textit{Resolve} field-of-view (FOV) and model it using a combination of a single ICM component and non-X-ray background (NXB) in \textsc{xspec}, as described in Section \ref{sec:methods}.
The Full-FOV spectrum is plotted in the upper panel of Figure \ref{fig:fullfov} along with the best-fit single-component model (black dashed line).
It is immediately apparent, however, that this simple model does not accurately describe the observed spectral features. 
Namely, the Ly$\alpha_2$ line is underpredicted by the single-component model (the Ly$\alpha_1/\alpha_2$ flux ratio is near unity), and there are significant residuals around the He$\alpha$ \textit{x}- and \textit{y}-lines. 

Given these residuals, we fit the spectrum with two ICM components. 
\setcounter{footnote}{0} 
This model improves the fit significantly, with a $\Delta$C-stat of $\sim 41$, corresponding to a Bayesian information criterion improvement of $\sim 70$\footnote{The likelihood of this improvement by random chance is $<10^{-7}$ \citep[see, e.g.,][]{churazov2015}.}.
This two-component model is also plotted in Figure \ref{fig:fullfov} as a solid black line.
This solution indicates that there are two velocity components (Figue \ref{fig:fullfov}, blue and red lines), at redshifts of $\sim 0.1664$ and $\sim 0.1706$. Both have velocity dispersions of $\sim 200$ km/s and are separated by a bulk velocity of $\sim 1030 \pm 80$ km/s (see Table \ref{tab:results}).
From this large bulk velocity separation, we can conclude that the ongoing merger in A1914 is primarily along the line of sight, rather than in the plane of sky.
Taken from the galaxy mean redshift $z=0.168$ \citep{barrena2013}, the first component is blueshifted by $\sim 420$ km/s, and the second is redshifted by $\sim 640$ km/s. 
Hereafter, we refer to these as the ``blue'' and ``red'' subclusters, respectively.

While exploring the parameter space of our two-component model, we also noticed a second local minimum, which has a smaller separation in bulk velocity between the components ($\sim 480 \pm 200$ km/s), a reduced velocity dispersion in the blue component ($\sim 100$ km/s), and an inflated velocity dispersion in the red component ($\sim 800$ km/s).
This extreme velocity dispersion is likely unphysical; however, this alternate solution is only disfavored by $\sim 0.3 \sigma$ {compared to the global minimum}, so cannot be ruled out statistically.
The global and local minima are labeled with a star and triangle in the bottom panel of Figure \ref{fig:fullfov}.
To fully explore the possible combinations of velocity dispersions and bulk velocity separation, we run an array of fits with fixed velocity dispersions $\sigma_\mathrm{red}$ and $\sigma_\mathrm{blue}$.
The best-fit bulk velocity separation ($\Delta V_{\rm bulk}$) between the two components of each combination of $\sigma$'s is plotted in the lower panel of Figure \ref{fig:fullfov}.
The global and local minima we identified are marked in this plot with a star and triangle, respectively.
We additionally plot 1/2/3 $\sigma$ constraints on $\sigma_\mathrm{red}$ and $\sigma_\mathrm{blue}$ as white contours.

In the best-fit result, the blueshifted component is around twice as bright within the FOV as the redshifted component.
This may indicate that the blue subcluster is more massive than the red; however, this is only inferred from the region covered by the Resolve FOV ($\sim500$ kpc) but not the entire cluster.
This situation is consistent with the galaxy redshift distribution, which is skewed blue (see Figure \ref{fig:mergergeometry} and \cite{barrena2013}).
{Assuming the gas velocity dispersion is due to isotropic turbulence,} the 3D turbulent Mach number ($M=\sqrt{3}\sigma/c_s$, where $c_s$ is the speed of sound) is $0.21_{-0.04}^{+0.04}$ and $0.23_{-0.07}^{+0.16}$ in the blue and red subclusters, respectively.
As we have distinguished the bulk velocity components along the line of sight, it is likely that the velocity dispersion does in fact reflect small-scale random motions.
These are similar to the Mach number of $\sim 0.24$ that was found in the core of another merging cluster; Coma \citep[][]{xrismcollab_coma}.
These velocity dispersions correspond to non-thermal pressure fractions $(M^2/(M^2+3/\gamma),$ where $\gamma=5/3)$ of $2.5_{-0.7}^{+0.8} \%$ and $4.0_{-2.8}^{+4.4} \%$, respectively. 
As we extracted this spectrum from the full \textit{Resolve} array, this method provides the statistically best measurements of the ICM parameters.
It does not, however, yield any information on the spatial structure of the cluster merger.

\subsection{Velocity mapping}
To investigate the spatial distribution of the two ICM components, we split the \textit{Resolve} array into nine regions and model these nine spectra with two ICM components each.
As in the previous fit, metallicities are tied between components.
As the statistics of each individual spectrum are very poor, due to the fine binning of the FOV, we additionally tie the temperatures, redshifts, and velocity dispersions of the nine red components to each other, and vice versa for the nine blue components.
The only free parameters in each region are the normalizations of each component.
This strategy essentially allows us to model the entire FOV with two ICM components, as in the previous experiment, while allowing the flux ratio between the red and blue subclusters ($F_\mathrm{red}/F_\mathrm{blue}$, where $F$ is the normalization of each ICM component) to vary across the nine subregions.
We find temperatures, bulk velocities, and velocity dispersions for the red and blue components that are broadly consistent with the best-fit results from the full FOV fitting (see Table \ref{tab:results}).

\begin{figure}
    \centering
    \includegraphics[width=\linewidth]{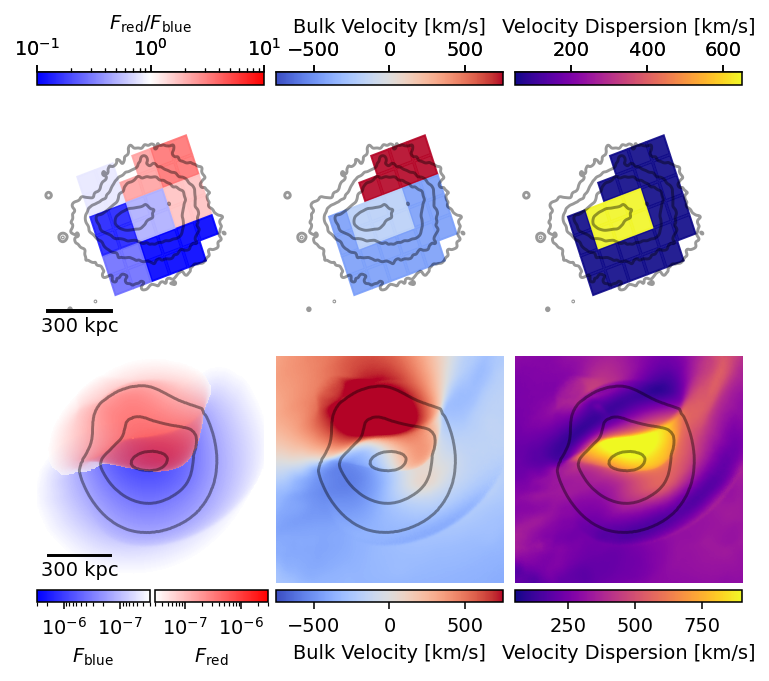}
    \caption{\textbf{Top:} Spatially resolved maps of the red/blue flux ratio (\textit{left}), bulk velocity (\textit{center}), and velocity dispersion (\textit{right}), derived from the nine-region and north-south-central binnings. Contours from the \chandra{} image of A1914 are shown in grey.
    \textbf{Bottom}: Simulated maps of X-ray fluxes from the red and blue subclusters (\textit{left}) and emissivity-weighted bulk velocity/velocity dispersion (\textit{center}/\textit{right}). Surface brightness contours are shown in black.
    }
    \label{fig:velocitymaps}
\end{figure}

The map of the flux ratio $F_\mathrm{red}/F_\mathrm{blue}$ is shown in Figure \ref{fig:velocitymaps}.
This map clearly indicates that the red subcluster dominates emission in the north/northwest of A1914, while the blue dominates the south/southeast.
Interestingly, the flux ratio map does not align with the cool, bright substructure seen in the \chandra{} image.
The fact that neither of the velocity components is coincident with this cool region rules out the Bullet-like merger scenario, i.e. this region is not a cool-core remnant passing through the cluster.
This region is actually spatially consistent with regions where the best-fit flux ratio is close to 1, implying that the two subclusters are ``overlapping'' in projection in this region.
This spatial distribution can be explained by a near-line-of-sight off-axis merger between the two subclusters. 
In this scenario, the red and blue subclusters are partially overlapping in projection, and their respective X-ray fluxes add up to create the bright region observed by \chandra{}.
We name this geometry a ``yin-yang merger'', because if the emission from each of these subclusters were separated, they would form a shape similar to a yin-yang symbol.

\begin{figure}
    \includegraphics[width=\textwidth]{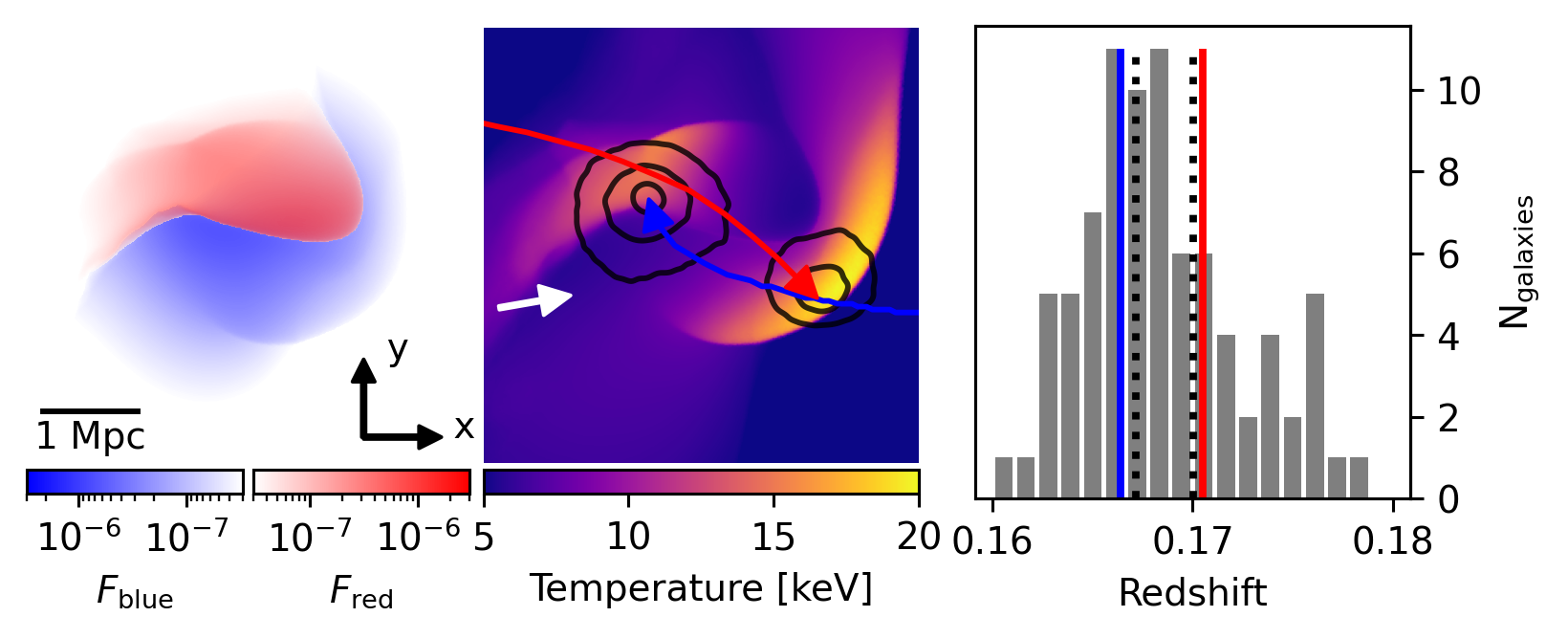}
    \caption{
    \textbf{Left:} Map of X-ray emission from the simulated red and blue subclusters, viewed perpendicular to the merger plane.
    \textbf{Center:} Simulated X-ray-weighted temperature map in the same projection. Here we additionally plot black contours showing the DM {mass distribution}, and red/blue arrows showing the trajectory of the corresponding subclusters during the merger. The white arrow indicates the projected line of sight presented in Figure \ref{fig:velocitymaps}. The line of sight is inclined to the x-y plane by $\simeq70^\circ$.
    \textbf{Right:} Redshift distribution of cluster member galaxies from \citet{barrena2013}, between z=0.16 and 0.18. Red and Blue ICM redshifts from \XRISM{} are shown as colored vertical lines. BCG redshifts are shown as dotted black lines.
    }
    \label{fig:mergergeometry}
\end{figure}

This merger scenario implies that there is a region of redshifted gas, a region of blueshifted gas, and a central region with an intermediate redshift.
Due to the overlapping bulk velocities, this central region should also have a significantly higher velocity dispersion if fitted with a single ICM component.
Using the flux ratio map as a guide, we split the FOV into three regions: a northern region dominated by redshifted gas, a southern region dominated by blueshifted gas, and a central region, corresponding to the bright substructure in the \chandra{} image.
We exclude the northeastern 2x2 pixel region as it does not clearly correlate with the bright structure or the blue/red subclusters.
We fit each of these regions with a single ICM component, which we map in Figure \ref{fig:velocitymaps}, while accounting for SSM (which ensures the large \XRISM{} PSF does not bias our measurement). 
As in the full-FOV fit, abundances are tied between the three regions, while temperature, redshift, and velocity dispersion are free.

As one can see in Figure \ref{fig:velocitymaps}, in the north and south regions we find redshifts (and bulk velocities) consistent with the red and blue subclusters found in the full FOV fit.
This is expected as, evidenced by the flux ratio map, emission in these regions is dominated by the red and blue subclusters, respectively.
In these regions we can only retrieve upper limits on the velocity dispersion.
In the central region, we find an intermediate redshift ($\sim 0.1673$) and high velocity dispersion (between $\sim 500$ and $800$ km/s).
These results are consistent with the proposed merger configuration, as the highest velocity dispersion is found in the central pixels, indicating the two velocity components are overlapping along the line of sight. 
We note that there is a degeneracy in this fit between the velocity dispersions in the central and south regions.
As we are correcting for SSM, it is possible to identify a local minimum where the high velocity dispersion comes from the southern region without significantly changing the model in either the central or southern regions. 
Completely ruling this solution out would require longer observations with \XRISM{}.

\begin{table}[]
    \centering
    \begin{tabular}{ccccccc}
    \input{figures/paramtable.txt}
    \end{tabular}
    \caption{Best-fit parameters and uncertainties from each ICM component in each binning strategy. Parameters marked with a dash are tied to the above component. V$_\mathrm{bulk}$ is calculated using the galaxy mean redshift 0.168. 
    }
    \label{tab:results}
\end{table}

\subsection{Comparison with tailored hydrodynamical simulations}
\label{sec:simulations}

To illustrate the proposed merger configuration, we perform hydrodynamical simulations of mergers between two idealized galaxy clusters using the moving-mesh code {\sc Arepo} \citep{Springel2010,Weinberger2020}. 
This type of simulation has been widely used to explore cluster merger processes \citep[e.g.,][]{Ricker2001,Poole2006,ZuHone2011}, particularly for modeling individual merging clusters \citep[e.g.,][]{Springel2007,Zhang2015,Bellomi2024}. 
In our simulations, each cluster consists of both gas and DM components, which are spherical and in equilibrium in their initial conditions.
Determining the initial condition of each cluster observationally is non-trivial.
Without loss of generality, we assume the initial gas and DM density radial profiles follow the Burkert and Navarro–Frenk–White (NFW) forms, respectively, as described in \citet{burkert1995,NFW1996}. 
We fix the concentration parameter ($=4$) for all our DM halos that determines the scale radius ($r_{\rm s}$) in the NFW model (see \citealt{Zhang2014} for more details of our numerical set-ups). The gas core radius in the Burkert profile is set as $r_{\rm s}/2$ and $2r_{\rm s}/3$ for the larger subcluster (blue) and the smaller subcluster (red), respectively. The slightly more extended gas distribution in the subcluster provides a better match with the observed X-ray surface brightness distribution. 
We fix the mass of the main cluster ($M_\mathrm{vir} = 8\times10^{14}\,M_\odot$, $V_\mathrm{vir}\approx1300$ km/s) in all our simulations, motivated by  weak-lensing measurements \citep{okabe2008}. 
We also fix the initial pairwise velocity ($V_0 = 900{\,\rm km\,s^{-1}}$) between the two merging clusters based on an empirical relation from cosmological simulations \citep{Dolag2013}, and explore other parameters that determine the merging halo trajectories, including the merger mass ratio ($\xi=2-10$) and impact parameter ($P_0=0.5-2\,{\rm Mpc}$). 
We include a passively advected scalar field that traces the gas initially associated with each subcluster.
We emphasize that our goal is to identify plausible merger scenarios for A1914 that generally follow the observed trends rather than to reproduce all observations in detail. 
Our experiments show that major mergers ($\xi=3$) with a moderate impact parameter ($P_0=1\,{\rm Mpc}$, which results in a pericenter distance of $\sim 200$ kpc) provide a good match to most of the observational features. 
We thus adopt this case for comparison with observations throughout this paper, unless stated otherwise. 

The lower panels of Figure \ref{fig:velocitymaps} show the simulated X-ray surface brightness of the blue and red subclusters, X-ray-weighted bulk velocity, and velocity dispersion.
Our simulation reproduces the main features seen in the XRISM velocity map and Chandra image, including the characteristic bulk velocity pattern, elongated X-ray excess in the central region and its corresponding high velocity dispersion (due to overlap of two halos). 
The X-ray map clearly shows the relative positions of the two merging subclusters in projection: the red subcluster dominates in the north, and the blue in the south. 
They overlap in projection, but have not yet mixed, producing the X-ray excess and high velocity dispersion described above.

To check that the proposed merger scenario accurately mimics observations from \chandra{} as well as the \XRISM{} results, we compare the projected X-ray surface brightness and temperature along the line of sight to the equivalent \chandra{} maps in Figure \ref{fig:chandra}.
We also compare the DM contours of the simulation to the weak lensing contours from \citet{mandal2019}.
It is readily apparent that the proposed merger geometry explains the northeast-southwest elongation of the mass distribution.
The X-ray surface brightness map is also a good match for the observations, showing a bright region, elongated along the east-west axis, in the center of the cluster.

The simulation shows an overall gas temperature of $\sim 9$ keV, consistent with the observed temperature map.
It also produces two shocks in the northeast and southwest which heat the gas to $\sim 11-13$ keV.
The positions of these shocks are consistent with those identified by \cite{rahaman2022}, {however the two southern shocks described in that work appear as a single, unbroken shock front in our simulation.
This difference could be explained if the shock front is disrupted by a galaxy or a pre-existing ICM substructure.}
The simulations also show a channel of cool gas separating the two shock-heated regions, however this channel is not fully formed in the \chandra{} temperature map.
{A cool region is clearly present to the east, though in the west there are hot substructures between the two shock-heated regions.
These hot structures are bordered to the west by several small eddy-like structures, which we label in Figure \ref{fig:chandra}.
It is possible these structures are remnants from a previous merger, or interaction between the ICM and the gas of the member galaxies. 
The small eddies could also be due to Kelvin-Helmholtz instabilities \citep[e.g.][]{nulsen1982}, however, these are not captured by our model, which aims to reproduce the overall trends in the gas and DM distribution.}
To demonstrate that A1914 fits the overall temperature structure proposed by the simulations, we extract and fit spectra from the regions shown as white boxes in Figure \ref{fig:chandra}.
With this method, we show that the gas between the two hotspots is $\sim 1$ keV cooler than the shock-heated regions, which is broadly consistent with the gas temperatures seen in simulations.

To have a complete view of the 3D geometry of the system, we also show maps of the X-ray surface brightness and gas temperature in a projection perpendicular to the merger plane in Figure \ref{fig:mergergeometry}.
DM density contours for the cluster are overlaid on the temperature map, along with the trajectories of the red and blue subclusters.
In this snapshot, the subclusters have passed their first closest approach, and their ICMs are wrapping around each other in a yin-yang shape.

One feature of A1914 that appears anomalous is the redshifts of the BCGs (0.167179 and 0.170012), which are moving slower with respect to the cluster mean ($\sim-210$ and $520$ km/s, respectively) than the gas components ($\sim -420$ and $640$ km/s).
Typically during a galaxy cluster merger, the collisionality of the ICM gas causes it and the DM component to decouple, such that the DM and galaxies are moving faster along the merger axis than the gas.
This can clearly be seen in the Bullet cluster \citep{markevitch2004}, where the DM and gas of the infalling subcluster have separated.
A1914 appears to show the opposite trend, though this can be explained by projection effects.
One can see from Figure \ref{fig:mergergeometry}, when the merger is viewed along the line of sight (denoted by the white arrow), the merging DM components have significant velocity in the plane of sky.
The interaction between the two subclusters has significantly altered the trajectory of the DM component, whereas the gas component is still moving primarily along the line of sight.
Indeed, our simulations predict the line-of-sight velocities of the blue and red DM potentials as $\sim -270$ and $750$ km/s, while the gas components have velocities of $\sim -500$ and $1000$ km/s, mirroring the trend seen in observations.

\section{Conclusions}
In this Letter, we presented the analysis of new \XRISM{} observations of {the merging cluster} A1914. 
From these observations, we measured the gas velocity structure of this cluster for the first time.
{Our main observational findings can be summarized as follows:}
\begin{itemize}
    \item
    Two bulk velocity components along the line-of-sight are required to accurately model the \XRISM{} spectrum of A1914, implying the cluster is undergoing a merger projected on the line-of-sight.
    \item 
    We recovered two ICM components moving with respect to each other at $\gtrsim 1000$ km/s. 
    These components are both composed of $\sim 9.5$ keV gas, with velocity dispersions of $\sim 200$ km/s (Figure \ref{fig:fullfov}.
    \item 
    An alternate solution with lower bulk velocities and much higher velocity dispersion ($\sim 800$ km/s) in one of the components exists, however this high velocity dispersion is not expected physically and this solution is slightly disfavored statistically.
    Further observations of A1914 are required to distinguish between these two possible solutions.
    \item 
    We map the flux ratio between these two components across the \textit{Resolve} FOV, and produce bulk velocity and velocity dispersion maps of the cluster (Figure \ref{fig:velocitymaps}). 
    These maps confirm the presence of two bulk velocity components that are offset in the plane of the sky, with the two components overlapping in projection {in the central region}.
    \item 
    These maps show {that} the central bright, cool substructure is not associated with either bulk velocity component, ruling out a head-on, Bullet-like merger scenario in this cluster.
    The velocity dispersion in this substructure is very high, indicating it arises from the overlap between the two bulk components.
\end{itemize}

We propose a ``yin-yang'' merger scenario to explain this velocity structure, where the two clusters are merging, mostly along the line of sight, with a high impact parameter. 
We support this scenario by comparing the observed velocity structure, X-ray morphology, ICM temperature map, and DM distribution to their simulated counterparts.
\begin{itemize}
    \item 
    We find the best match to observations from a simulations with merger mass ratios $\xi\sim 3$ and an impact parameter of $\sim$ 1 Mpc.
    \item
    The bulk velocity and velocity dispersion maps of the simulated merger {with these parameters} are consistent with the measured gas velocities.
    \item 
    This merger scenario also resolves the apparent tension between an E-W elongation in the X-ray morphology, caused by the overlapping subclusters, and the NE-SW elongation of the DM distribution.
    \item
    The velocities of the BCGs, which are lower than the bulk velocities of the gas, are also reproduced in this simulation.
    Our merger scenario implies these galaxies and their corresponding DM halos have changed trajectory significantly through gravitational interaction, and should therefore have a significant plane-of-sky velocity.
    \item 
    The large-scale temperature structure and X-ray morphology of A1914 is reproduced in our simulation, however some substructures exist in A1914 that are not present in the simulation (Figure \ref{fig:chandra}).
    It is likely that these substructures are leftover from a previous merger, or due to interaction between the ICM and cooler gas in galaxies.
    The presence of eddies in the X-ray image of A1914 associated with the substructures support this idea.
    
\end{itemize}

This Letter demonstrates the power of \XRISM{}'s spectral resolution to {accurately} measure gas velocities {in the ICM}, disentangling line-of-sight cluster mergers and allowing for better reconstruction of their geometries.
{Improving our understanding of this type of merger will help to complete our picture of galaxy clusters as sites of hierarchical structure formation and heating in the universe.}

\begin{acknowledgments}
AH would like to thank Henk Hoekstra for providing the observed weak-lensing contours shown in Figure \ref{fig:chandra}. 
AH was partially supported by NASA grant 80NSSC25K7693.
AH, CZ, and IZ acknowledge partial support from the Alfred P. Sloan Foundation through the Sloan Research Fellowship. 
CZ and IZ were partially supported by NASA grant 80NSSC18K1684. 
CZ was supported by the GACR grant c. 
WF acknowledges support from from the Smithsonian Institution, the Chandra High Resolution Camera Project through NASA contract NAS8-03060, and NASA Grants 80NSSC19K0116 and GO1-22132X. 
This work was completed in part with resources provided by the University of Chicago’s Research Computing Center. 
\end{acknowledgments}



%
\facilities{XRISM (Resolve), CXO.}


\appendix
\section{Predicted kinetic SZ signal from A1914}

\begin{figure}
    \includegraphics[width=\textwidth]{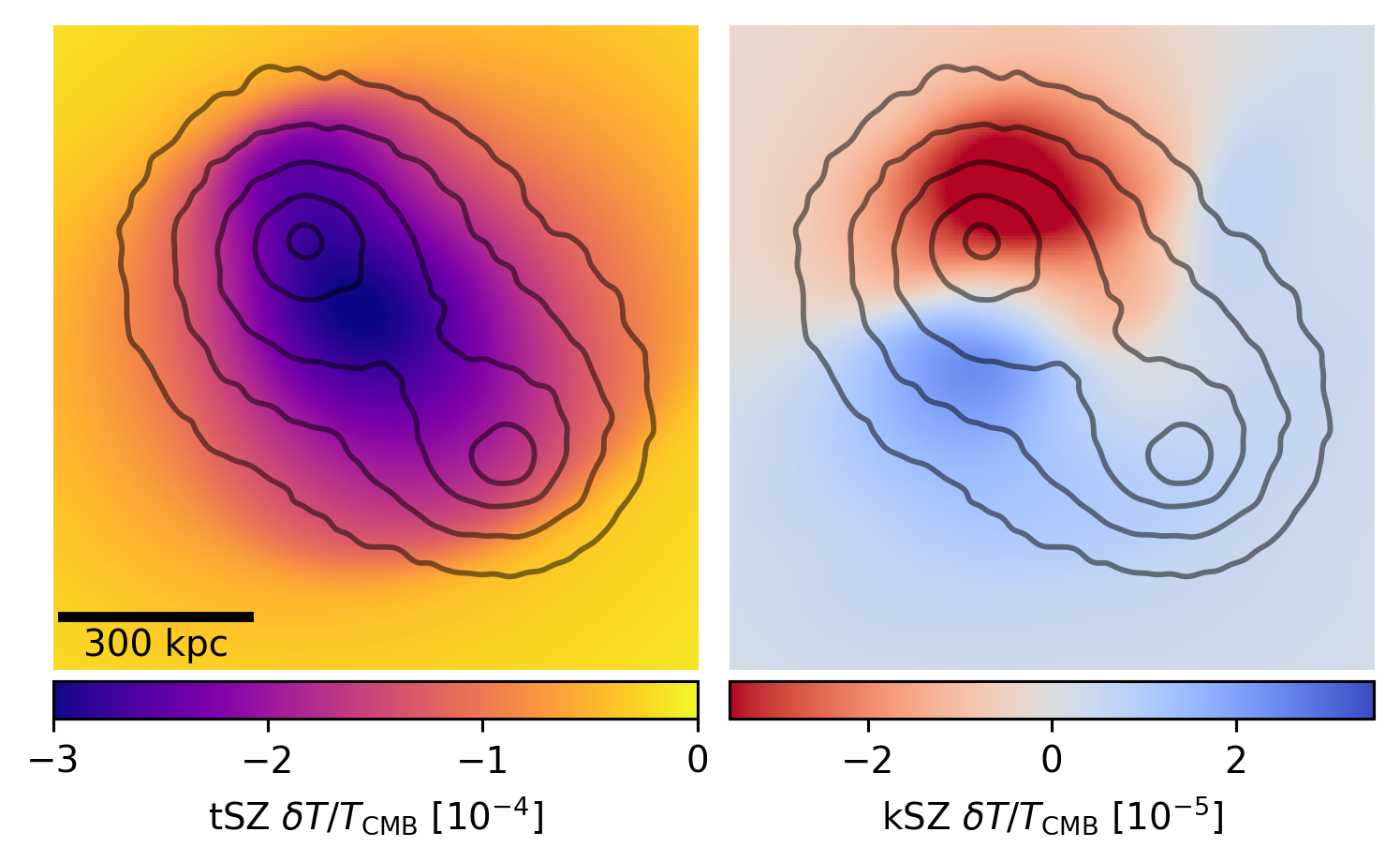}
    \caption{
    Predicted distributions of the tSZ (left) and kSZ (right) effects at 150 GHz, in LOS projection equivalent to Figure \ref{fig:chandra}. Both maps have been smoothed with a 8 arcsecond Gaussian. Black contours show DM matter distribution.
    } 
    \label{fig:SZeffect}
\end{figure}

The kinetic Sunyaev-Zel'dovich effect (kSZ) has been detected in a handful of galaxy clusters \citep{sayers2013,adam2017,sayers2019,silich2024}. 
Since A1914 is undergoing a nearly line-of-sight merger, we expect to observer a strong gradient in the kSZ signal across the center of the cluster, making it an interesting target for future kSZ observations.
Figure ~\ref{fig:SZeffect} presents predictions of the system’s SZ effect based on our simulations. 
The left and right panels show the relative changes in the cosmic microwave background (CMB) temperature  caused by the thermal SZ (tSZ) and kSZ effect at a frequency $150\,$GHz, respectively. 
Relativistic corrections are included in the estimation \citep{Itoh1998,Nozawa2000}. 
Both maps are smoothed with an 8 arcsecond Gaussian to approximate the beam size of the NIKA2 instrument \citep{ruppin2018}. 
The kSZ peaks at $\sim -4$ and $2.3 \cdot 10^{-5}$ in the red and blue components, approximately $10\%$ of the tSZ peak strength.
\\

\bibliography{references}{}
\bibliographystyle{aasjournalv7}

\end{document}

%% file: figures/paramtable.txt
Source & kT [keV] & Fe & Ni & z & V$_\mathrm{bulk}$ [km/s] & $\sigma$ [km/s] \\
\hline
\\
\multicolumn{7}{c}{Full-FOV Fit, 1-Component} \\
\multicolumn{7}{c}{C-stat=2082, 2121 DoF} \\
\hline
ICM & ${9.5}_{-0.2}^{+0.2}$ & ${0.27}_{-0.02}^{+0.02}$ & ${0.26}_{-0.17}^{+0.18}$ & ${0.1665}_{-0.0001}^{+0.0001}$ & ${-375}_{-29}^{+38}$ & ${257}_{-36}^{+61}$ \\
\\
\multicolumn{7}{c}{Full-FOV Fit, 2-Component} \\
\multicolumn{7}{c}{C-stat=2041, 2117 DoF} \\
\hline
Blue & ${9.5}_{-0.5}^{+0.4}$ & ${0.31}_{-0.02}^{+0.02}$ & ${0.25}_{-0.18}^{+0.19}$ & ${0.1664}_{-0.0001}^{+0.0001}$ & ${-416}_{-27}^{+27}$ & ${190}_{-30}^{+30}$ \\
Red & ${9.8}_{-0.9}^{+1.3}$ & -- & -- & ${0.1705}_{-0.0004}^{+0.0002}$ & ${641}_{-100}^{+53}$ & ${245}_{-94}^{+113}$ \\
\\
\multicolumn{7}{c}{Nine-Region Fit} \\
\multicolumn{7}{c}{C-stat=9087, 10704 DoF} \\
\hline
Blue & ${9.5}_{-0.3}^{+0.3}$ & ${0.31}_{-0.02}^{+0.02}$ & ${0.36}_{-0.19}^{+0.19}$ & ${0.1663}_{-0.0001}^{+0.0002}$ & ${-436}_{-22}^{+41}$ & ${173}_{-26}^{+32}$ \\
Red & ${9.7}_{-0.4}^{+0.4}$ & -- & -- & ${0.1703}_{-0.0004}^{+0.0004}$ & ${595}_{-99}^{+107}$ & ${436}_{-82}^{+128}$ \\
\\
\multicolumn{7}{c}{North-South-Central Fit} \\
\multicolumn{7}{c}{C-stat=4859, 5051 DoF} \\
\hline
North & ${12.0}_{-1.2}^{+1.1}$ & ${0.33}_{-0.02}^{+0.02}$ & ${0.22}_{-0.19}^{+0.20}$ & ${0.1709}_{-0.0004}^{+0.0002}$ & ${751}_{-90}^{+41}$ & $< 162$ \\
South & ${9.7}_{-0.4}^{+0.4}$ & -- & -- & ${0.1663}_{-0.0001}^{+0.0001}$ & ${-448}_{-26}^{+27}$ & $< 110$ \\
Central & ${8.9}_{-0.3}^{+0.4}$ & -- & -- & ${0.1673}_{-0.0004}^{+0.0004}$ & ${-176}_{-112}^{+95}$ & ${644}_{-110}^{+123}$ \\